# Colors of Extreme Exo-Earth Environments

Siddharth Hegde[1,*] and Lisa Kaltenegger[1,2]

[1]Max Planck Institute for Astronomy, Heidelberg, Germany;

[2]Harvard-Smithsonian Center for Astrophysics, Cambridge, Massachusetts, USA.

[*]E-mail: hegde@mpia.de; sidworld@gmail.com

Abstract

The search for extrasolar planets has already detected rocky planets and several planetary candidates with minimum masses that are consistent with rocky planets in the habitable zone of their host stars. A low-resolution spectrum in the form of a color-color diagram of an exoplanet is likely to be one of the first post-detection quantities to be measured for the case of direct detection.

In this paper, we explore potentially detectable surface features on rocky exoplanets and their connection to, and importance as, a habitat for extremophiles, as known on Earth. Extremophiles provide us with the minimum known envelope of environmental limits for life on our planet.

The color of a planet reveals information on its properties, especially for surface features of rocky planets with clear atmospheres. We use filter photometry in the visible waveband as a first step in the characterization of rocky exoplanets to prioritize targets for follow-up spectroscopy.

Many surface environments on Earth have characteristic albedos and occupy a different color space in the visible waveband (0.4-0.9 µm) that can be distinguished remotely. These detectable surface features can be linked to the extreme niches that support extremophiles on Earth and provide a link between geomicrobiology and observational astronomy. This paper explores how filter photometry can serve as a first step in characterizing Earth-like exoplanets for an aerobic as well as an anaerobic atmosphere, thereby prioritizing targets to search for atmospheric biosignatures.

1. Introduction

Several rocky (e.g., Léger et al., 2009; Batalha et al., 2011) as well as potentially rocky (e.g., Mayor et al., 2009, 2011) exoplanets have been detected. Three super-Earths, consistent with rocky planetary models, Gliese 581d, HD 85512b, and Gliese 667Cc, orbit their host stars within the habitable zone (Udry et al., 2007; Bonfils et al., 2011; Pepe et al., 2011; Anglada-Escudé et al., 2012). In addition, NASA's Kepler mission has recently announced several potentially rocky exoplanet candidates (see Batalha et al., 2012; Borucki et al., 2011, 2012) in the habitable zone.

A comprehensive suite of tools will be needed to characterize such planets since a mere detection of a rocky planet in the habitable zone does not guarantee the planet to be habitable (e.g., Selsis et al., 2007; Kaltenegger and Sasselov, 2011). For the case of direct detection of exoplanets, detailed characterization in regard to the habitability of Earth-like extrasolar planets or "exo-Earths" is achieved by studying the atmospheric and surface properties of the planet in contention. Filter photometry is a tool with which to initially characterize exoplanets. A color-color diagram distinguishes giant planets from rocky ones for Solar System objects (see Traub, 2003b; Crow et al., 2011, for details).

The position of an extrasolar planet in a color-color plot can in principle show analogies to the Solar System, approximate the basic physical properties of a planet (Traub, 2003a, 2003b), and place constraints on its atmospheric composition (Crow et al., 2011). Exploring surface features of Earth-like planets becomes possible if either no significant cloud cover exists on an exoplanet or the signal-to-noise ratio (SNR) of each observation is sufficiently high to remove the cloud contribution from the overall detected signal (e.g., Ford et al., 2001; Pallé et al., 2008; Cowan et al., 2009, 2011; Fujii et al., 2011).

In this paper, we focus on Earth-like planets within this framework and especially on remote differentiation of the different environments found on Earth that are known to support extreme forms of life or "extremophiles." Small changes in temperature, pH, or other physical and geochemical factors (see section 2) can lead to such environments being dominant on a potentially habitable exoplanet, which could govern evolution of life. These various "extreme" surface environments on Earth have characteristic albedos in the visible waveband (0.4-0.9 µm) that could be distinguished remotely. We therefore explore the color signatures that are obtained from the surface environments inhabited by extremophiles as well as test our approach by using measured reflection spectra of extremophiles.

Note that detection of such surface features of environments in a reflection spectrum alone is not a reliable detection of life on an exoplanet. This diagnostic can only be used in combination with atmospheric properties (see, e.g., Cockell et al., 2009).

Several groups have focused their attention on the vegetation red edge (VRE) (also known as the chlorophyll signature) to detect direct signatures of life remotely (e.g., Arnold et al., 2002; Woolf et al., 2002; Seager et al., 2005; Montañés-Rodríguez et al., 2006; Tinetti et al., 2006; Kiang et al., 2007a, 2007b). The VRE is a surface feature of terrestrial land plants that has been widespread since about 0.46 billion years ago (Carroll, 2001; Igamberdiev and Lea, 2006; Kiang et al., 2007a) and is characterized by a strong increase in the reflectivity at near-IR wavelength regions longward of 0.75 µm. In this paper, we expand that approach for different life-forms by including surfaces that provide environmental conditions for extremophiles on Earth over geological times. We consider different classes of extremophiles that have adapted to severe physical or geochemical extremes in order to explore the known limits for habitability on exoplanets. We thereby link geomicrobiology to observational astronomy by exploring the low-resolution characterization of an Earth analog planet for different surface environments and the known extreme forms of life on Earth that such environments could harbor.

In this work, Section 2 discusses the different types of extremophiles on Earth. Section 3 focuses on surface characteristics of different environments that support extremophiles and links those extreme environments to remotely detectable observables. Section 4 presents our results with a low-resolution color-color diagram to distinguish the different environments, explores the effect of mixed surfaces, and builds a link to extremophiles for aerobic and anaerobic atmospheres. We discuss our results in Section 5 and conclude in Section 6.

## 2. Extremophiles on Earth

The search for life on extrasolar planets relies on defining limits for life in regard to its evolution and distribution as observed on Earth. These limiting factors in turn correspond to physical or chemical parameters, which act as templates while looking for life elsewhere. To provide a wide definition, we therefore focus on organisms that live under extreme conditions on Earth. Extremophiles (literally "lovers of extreme environments") are organisms that live and thrive in very harsh environmental conditions. These environmental "extremes" are defined in terms of physical (such as temperature, radiation, pressure, etc.) and geochemical (desiccation, salinity, pH, etc.) extremes (see, e.g., Rothschild and Mancinelli, 2001). Observations on Earth indicate that life is ubiquitous even in extreme niches as long as there is liquid water, an energy source for metabolism, and a source of nutrients that helps in building and maintaining cellular structures (Rothschild, 2009).

We explore the parameter space of extremophiles to look for signatures of extraterranean life by focusing on extreme environments that those organisms inhabit on Earth.

Table 1 shows that multiple organisms have evolved to function at different environmental extremes (e.g., temperature, pH), which suggests that such evolutionary adaptation is not a singular event (Rothschild, 2008) and carbon-based life-forms may evolve in similar environments on extrasolar rocky planets. Regarding temperature extremes, extremophiles are categorized as hyperthermophiles (temperature for growth is >80°C) that are, for example, isolated from submarine hydrothermal vents, and psychrophiles (temperature for growth is <15°C) that are, for example, isolated from glaciers. Alkaliphiles grow at a pH >9 and are found, for example, in soda lakes; acidophiles grow at pH <5 and are, for example, found in acid mine drainages. Piezophiles are organisms that thrive under extreme pressure conditions such as the Mariana Trench, which has a hydrostatic pressure of ~1,100 bars (Marion et al., 2003). Halophiles grow under high salt concentrations and can be found, for example, in salt lakes. Xerophiles grow with very little water and reside, for example, in sand deserts, ice deserts, and salt flats like the Atacama Desert in Chile. Endoliths live inside rocks such as sandstone, which protect the organisms by attenuating the UV radiation while allowing the photosynthetically active radiation through its upper translucent surface (Southam et al., 2007) and thereby allowing for photosynthetic metabolism. This phenomenon is often described as "cryptic photosynthesis" due to the effective shielding of any specific reflection signature that could indicate the biota by the overlying rock surface in the reflection spectrum. Rocks protect the organisms residing within against low temperatures, UV radiation, and severe desiccation (Cockell et al., 2009; Canganella and Wiegel, 2011). Endolithic communities are also found in complete darkness, whereby they receive their energy by reducing sulfate and iron among other metals found in the host rock (Cavicchioli, 2002).

Limits as stated in Table 1 for the different classes of extremophiles are based on activity and not mere survival, except for radiation. Radiation-tolerant extremophiles survive high doses of ionizing and UV radiation but do not grow optimally under those conditions, as is seen in laboratory experiments (Venkateswaran et al., 2000; Rothschild and Mancinelli, 2001). Exceptional levels of ionizing and UV radiation rarely occur naturally on present-day Earth, and the tolerance of extremophiles to radiation is suggested to be a by-product of their resistance to desiccation (Mattimore and Battista, 1996; Battista, 1997; Ferreira et al., 1999).

## 3. Detectable surface features

Table 1 shows that most extremophiles live in subsurface conditions either as a measure to protect themselves or a means to gain access to the required nutrients provided by the host environment. Therefore, when observing surface features remotely, unlike surface vegetation, one does not generally detect extremophiles in a reflection spectrum unless the extremophilic organism is living close to or on top of the surface environment. In this paper, we focus on the surface features of environments that support extremophiles residing within them. To complement the approach, we also include the albedos of extremophiles in our study when available, to show the applicability to surface reflection of extremophiles as well. Table 1 shows the extreme environment or source that supports the various classes of extremophiles and the surface features that are detectable remotely. The data for these spectra were obtained from the ASTER spectral library (Baldridge et al., 2009) and the USGS digital spectral library (Clark et al., 2007).

Table 1

| Environmental Parameter | Class | Defining Growth Condition | Environment / Source | Remotely Detectable Observable | Example Organisms |
|---|---|---|---|---|---|
| High Temperature | Hyperthermophile | > 80 °C | Submarine Hydrothermal Vents | Water | *Pyrolobus fumarii, Strain 121* |
|  | Thermophile | 60 - 80 °C | Hot Spring |  | *Synechococcus lividis, Sulfolobus* sp. |
| Low Temperature | Psychrophile | < 15 °C | Ice, Snow | Ice, Snow | *Psychrobacter, Methanogenium* spp. |
| High pH | Alkaliphile | ph > 9 | Soda Lakes | Salt | *Bacillus firmus OF4, Haloanaerobium alcaliphilum* |
| Low pH | Acidophile | ph < 5 (typically much less) | Acid Mine Drainage, Volcanic Springs | Acid Mine Drainage | *Picrophilus oshimae/torridus, Stygiolobus azoricus* |
| High Pressure | Piezophile | High pressure | Deep Ocean eg. Mariana Trench | Water | *M.kandleri, Pyrococcus* sp., *Colwellia* sp. |
| Radiation | - | Tolerates high levels of radiation | Sunlight eg. High UV radiation | Sand, Rocks | *Deinococcus radiodurans, Thermococcus gammatolerans* |
| Salinity | Halophile | 2 - 5 M NaCl | Salt Lakes, Salt Mines | Salt | *Halobacteriaceae, Dunaliella salina, Halanaerobacter* sp. |
| Desiccation | Xerophile | Anhydrobiotic | Desert, Rock surfaces | Sand, Rocks | *Artemia salina, Deinococcus* sp., Lichens, *Methanosarcina barkeri* |
| Rock-dwelling | Endolith | Resident in rock | Upper subsurface to deep subterranean | Rocks | Lichens, Cyanobacteria, *Desulfovibrio cavernae* |

TABLE 1: Classification of Extremophiles.

"Remotely Detectable Observable" denotes the surface reflection signatures that can be observed remotely for the extreme environments considered here, which support different classes of extremophiles. Adapted from Rothschild and Mancinelli (2001), Cavicchioli (2002), Marion et al. (2003), Pikuta et al. (2007), Rothschild (2009), Canganella and Wiegel (2011).

The direct albedos of the extremophiles obtained from the spectral libraries were limited to a sample of three organisms that is composed of lichens, bacterial mats, and red algae in acid mine drainage (AMD) and have been included in this work. Lichens are composite organisms that consist of a fungus with a photosynthetic partner, usually either a green alga or cyanobacterium (Marion et al., 2003). Lichens are desiccation-resistant and occur at some of the most extreme environments on Earth, such as hot deserts, but are also found on top of rock surfaces. The bacterial mat spectrum used here is composed of two thermophilic species, the photosynthetic bacterium *Chloroflexus aurantiacus* and the cyanobacterium *Synechococcus lividus*, which are found at Octopus Springs in Yellowstone National Park, USA. This microbial mat forms at a temperature of ~65°C (Rothschild and Mancinelli, 2001), which is cool compared to the temperature regimes occupied by the hyperthermophilic organisms in hydrothermal vents [>110°C (Marion et al., 2003)]. The lower temperature still permits photosynthesis, as chlorophyll breaks down above ~75°C (Rothschild and Mancinelli, 2001). Hence, these organisms grow very close to the water's surface such that they receive enough sunlight to carry out photosynthesis; therefore, the reflection spectrum of the microbial mat will have a major contribution to the overall albedo [measured on site with a field spectrometer (Clark et al., 2007)]. The spectrum of red algae in AMD is one where red algae are coating rocks at ~10 cm depth below the surface.

Some surfaces, such as water, can be linked to several extremophiles like hyperthermophiles in submarine hydrothermal vents as well as piezophiles found deep in the ocean. Snow can be linked to psychrophiles living in cold environments in this model. Soda lake, a type of salt lake with a high content of sodium salts, in particular, chlorides or sulfates, can be linked to both alkaliphiles as well as halophiles. AMD links to Acidophiles. Sand is used here as the surface feature for xerophiles and radiation-resistant extremophiles. Finally, exposed rocks like limestone link to endoliths that live inside the rock as well as xerophiles and radiation-resistant organisms, which live on the rock surface. Note that some surface features like water are not a unique indicator for extreme environments and can also indicate environmental conditions that are not extreme. Our aim in this work is to explore a wide range of surfaces that can support extreme forms of life. But the reverse, that these surfaces all need to support life, is not given. By considering the range of extreme environments, we also include non-extreme forms of life in our model. Information on a planet's potential habitability can only be obtained once atmospheric properties and biosignatures are detected in the planet's atmosphere.

The results obtained in this work aim to prioritize targets for spectroscopic characterization. The method presented here provides a first characterization for prioritizing exoplanet targets for follow-up spectroscopy.

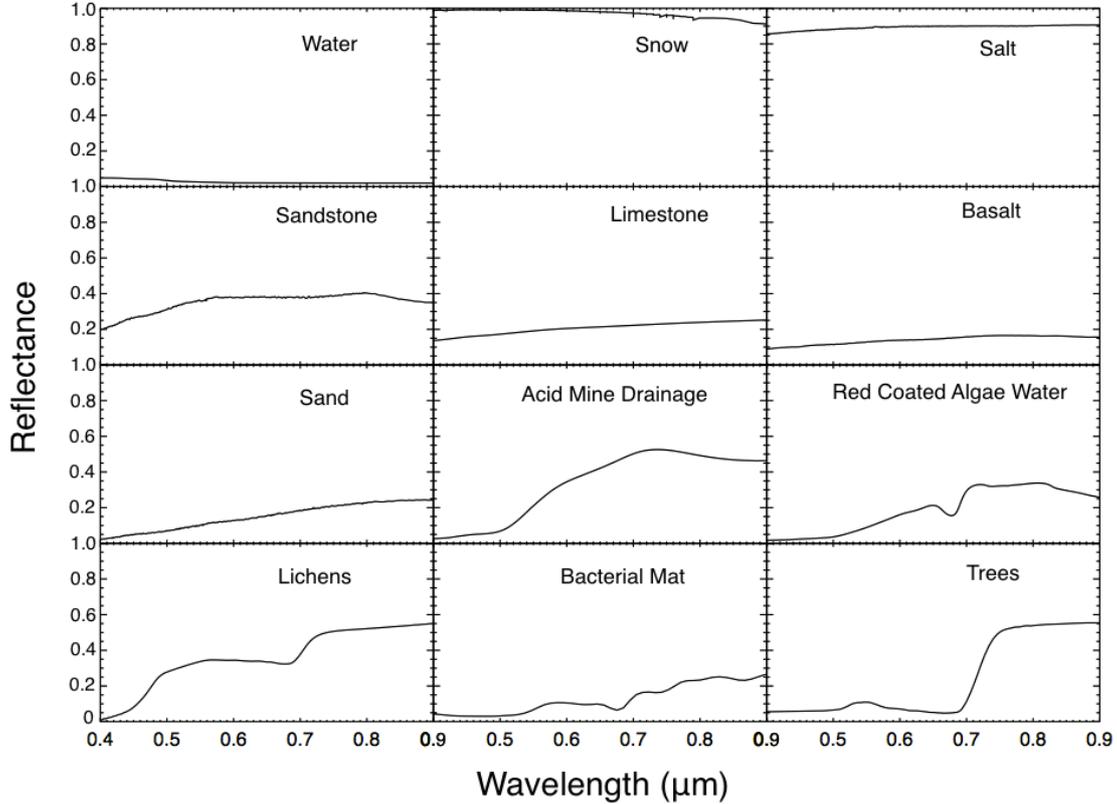

FIG. 1: Characteristic reflection spectra of different surfaces on present-day Earth that are known to support extremophiles as well as bacterial mat[1], lichens, and trees[2].

Fig. 1 shows the corresponding surface albedo for water, snow, salt, sandstone, limestone, basalt, sand, AMD, red coated algae water, lichens, bacterial mat, and trees. Apart from the remotely detectable surface features for extreme environments, we also include trees (deciduous) for vegetation (VRE signature), as a reference to other studies for comparison. Other terrestrial land plants occupy a similar position in our color-color diagram.

---

[1] The bacterial mat spectrum used here is composed of two thermophilic species, the photosynthetic bacterium *Chloroflexus aurantiacus* and the cyanobacterium *Synechococcus lividus*.

[2] Data obtained from the ASTER spectral library (Baldridge et al., 2009) and the USGS digital spectral library (Clark et al., 2007).

## 4. Results

Using the albedos of extremophiles as well as different surfaces they reside in as shown in Fig. 1, we explore the low-resolution characterization of rocky Earth-like extrasolar planets in the visible waveband (0.4-0.9 µm) by making use of a color-color diagram. Note that here we inherently assume an Earth analog atmosphere and suppression of the stellar light to make this comparison (see Discussion). To assess general detectability, we first assume full surface coverage of a particular surface (e.g., water, snow). Using these approximations, we use a color-color diagram to distinguish planetary environments remotely.

Color is the difference of magnitudes in two filter bands and is defined as:

$$C_{AB} = A - B = -2.5 \log_{10}\left(\frac{r_A}{r_B}\right) \quad (1)$$

where, $r_A$ is the reflectivity in band A and $r_B$ in band B.

We use the standard Johnson-Cousins BVI broadband filters, blue (B) = 0.4-0.5 µm, visible (V) = 0.5-0.7 µm, near-IR (I) = 0.7-0.9 µm, here. In principle, any number of filter bands can be used as long as the bandpass definitions are larger than the expected measurement noise and there is a high enough signal available per filter, which would allow finer distinctions.

Fig. 2 shows the B-V versus B-I color-color diagram, which distinguishes the environments (shown in Fig. 1) clearly. Surfaces with high reflectivities—snow, salt, and rocks that support endoliths, such as sandstone, limestone, and basalt—group together, as indicated by color coding in Fig. 2. Trees are shown as a reference to other VRE studies. Trees do not group with the other photosynthesis-based species considered here that contain chlorophyll pigments like the *Synechococcus*-bearing bacterial mats or lichens (see Discussion). AMD and red coated algae water (i.e., AMD water with red alga living at ~10 cm depth below the surface) group together.

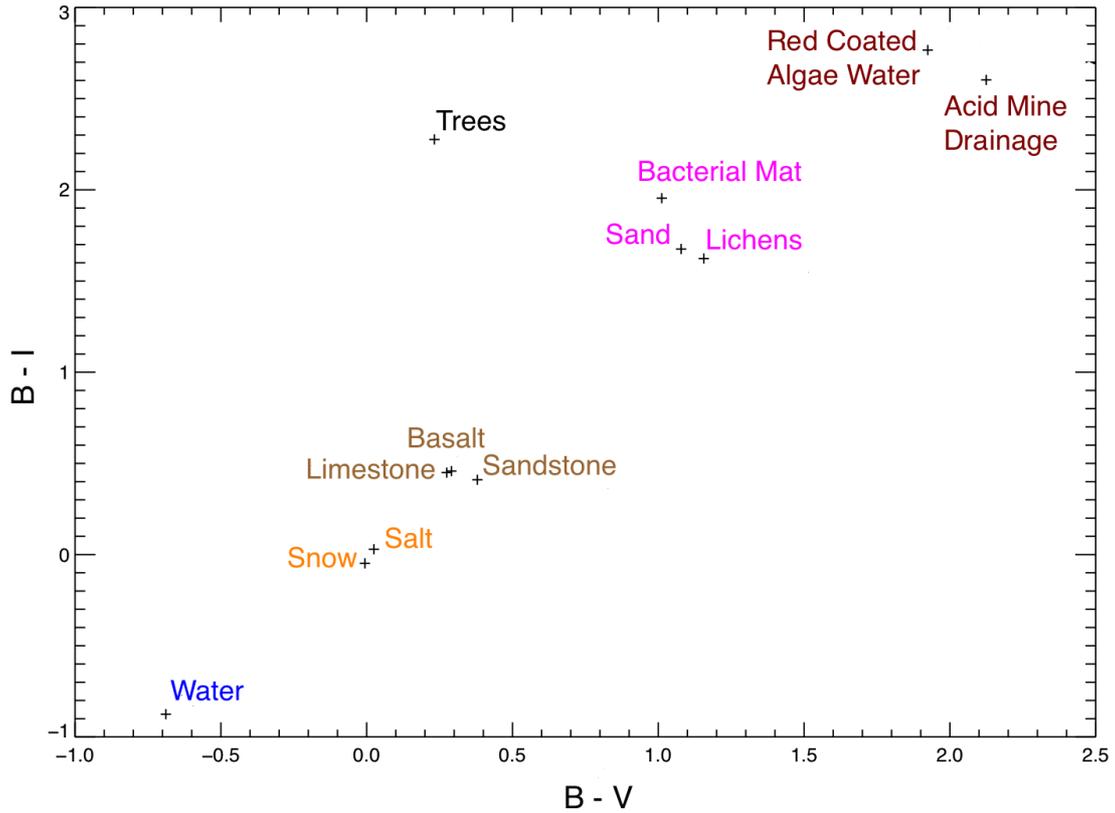

FIG. 2: Color-color diagram based on observed reflection spectra of characteristic surfaces that support extremophiles on Earth as well as bacterial mat, lichens, and trees (using conventional Johnson-Cousins BVI filters). Trees are shown here in black as reference to other VRE studies.

Our results indicate that different surface types can be distinguished in a color-color diagram. Highly reflective surfaces like snow, salt, and the different rocks form two distinct groups. AMD and red algae at ~10 cm depth in AMD group together, which suggests that extremophiles living in subsurface conditions can be linked to remotely detectable surface features.

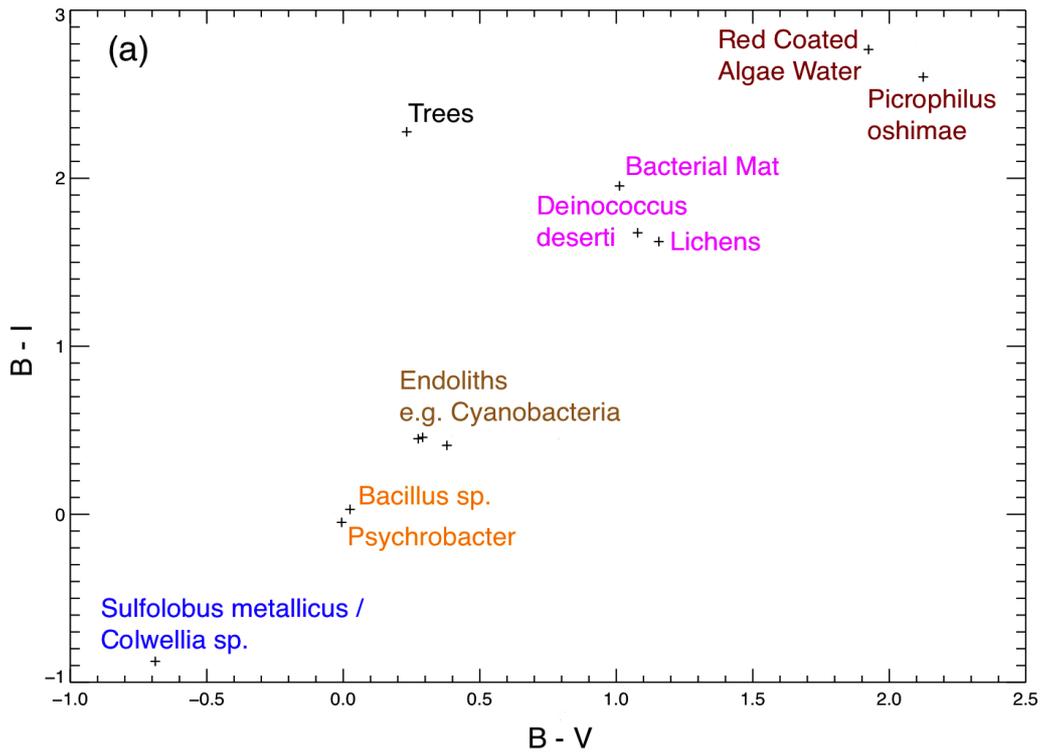

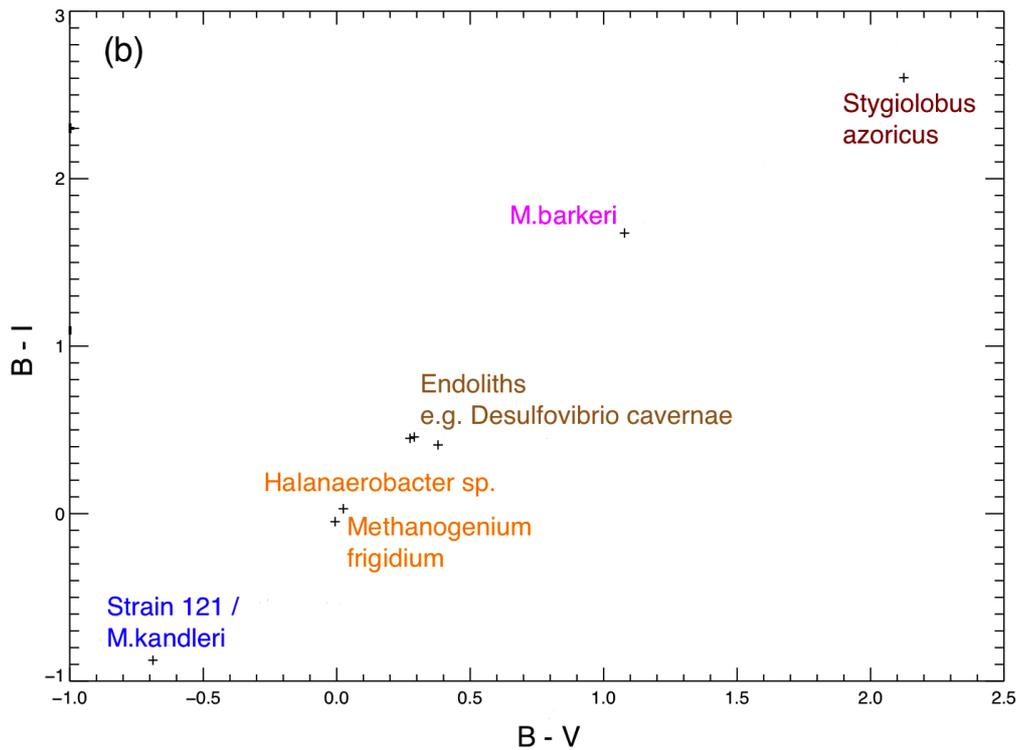

FIG. 3: Color-color diagram based on observed reflection spectra of characteristic surfaces that support extremophiles as well as bacterial mat and lichens in (a) an aerobic atmosphere (top) and (b) an anaerobic atmosphere (bottom).

Fig. 3 links the remotely detectable surface features to habitability with the use of different kinds of extremophiles that exist under extreme environments on Earth. On the basis of geological evidence, we know that Earth has been anaerobic for parts of the history of life. Significant levels of free $O_2$ first appeared in the biosphere around 2.4 billion years ago (Cloud, 1972; Holland, 1994, 2006; Pavlov et al., 2001). Life on present-day Earth inhabits environments that range from those that exhibit strictly anaerobic (obligate anaerobes) conditions to those that exhibit strictly aerobic (obligate aerobes) conditions. Fig. 3 indicates where the different classes of extremophiles fall, for an environment that is aerobic (3a) or anaerobic (3b). The VRE surface albedo is shown for reference. Note that the data points for photosynthesis-based organisms have been removed from the anaerobic plot, as they are aerobic in nature.

Our model initially assumes full surface coverage of one particular surface in order to explore the general detectability of surface effects in a color-color diagram. This assumption is valid if changes in temperature, pH, or other physical or geochemical parameters differ slightly from Earth's. Therefore a particular environment that is considered extreme on current Earth could dominate a potentially habitable rocky planet and thereby govern the available environment for life. Fig. 2 shows that the surface effects can be detected remotely.

Based on our current understanding, liquid water is one of the main ingredients necessary for life. Water is also the dominant surface fraction on Earth, but its surface coverage on exoplanets is generally not known. Therefore, we explore the parameter space by investigating the effect of water as a second surface on the detectability of different extreme surface environments on a hypothetical exoplanet. Note that one could allow for a wide range of potential combinations of surface types in this parameter space. With water being the necessary ingredient for life, we constrain the number of additional surfaces to water in our analysis (see, e.g., Fujii et al., 2011, for a detailed analysis on the retrieval of different surface features on present-day Earth). Following our initial argument of slight physical or chemical changes on a habitable rocky planet, one surface environment should dominate our extreme exo-Earth model.

To explore the parameter space of the fraction of water on an exoplanetary surface, we calculate a set of color-color diagrams with varying water surface fraction to explore its effect on prioritization of exoplanet targets for follow-up spectroscopy (see Fig. 4). The data point represented in blue denotes the position of present-day

Earth: it is modeled by assigning 70% of the planetary surface as ocean, 2% as coast, and 28% as land. The land fraction consists of 60% vegetation, 9% granite, 9% basalt, 15% snow, and 7% sand (following Kaltenegger et al., 2007).

Fig. 4 shows that the addition of water surface fraction from 10% to 90% moves the position of the planet in the color-color diagram along a diagonal. We define the region encompassing planets dominated by extreme environments with differing surface fractions of water between contours denoted by region I "extreme Earths." Note that some surface features like water are not a unique indicator for extreme environments and also indicate environmental conditions that are not extreme. Allowing for non-extreme habitable environments, like vegetation, widens the contours of extreme Earths in the color-color diagram (denoted region II) to a wider contour of "habitable planets."

## 5. Discussion

The method presented in this paper provides a strategy with which to prioritize targets for follow-up spectroscopy once rocky extrasolar planets have been identified either by their physical properties or with the use of a color-color diagram such as those used to distinguish giant and rocky planets in our Solar System (see Traub, 2003b; Crow et al., 2011). Our aim in this work is to explore a wide range of surfaces that can support extreme forms of life. But the reverse, that these surfaces all need to support life, is not given.

As a first-order approximation, we assume here that the planet has a see-through atmosphere such that clouds and hazes do not obscure the surface signatures of the planet. Exploring surface features of exoplanets is only possible if either no significant cloud cover exists on an exoplanet or the SNR of each observation is sufficiently high to remove the cloud contribution from the overall detected signal (see Pallé et al., 2008). This holds true for directly observable surface reflection features of life like the VRE from terrestrial land plants and for the direct and indirect features shown in this paper.

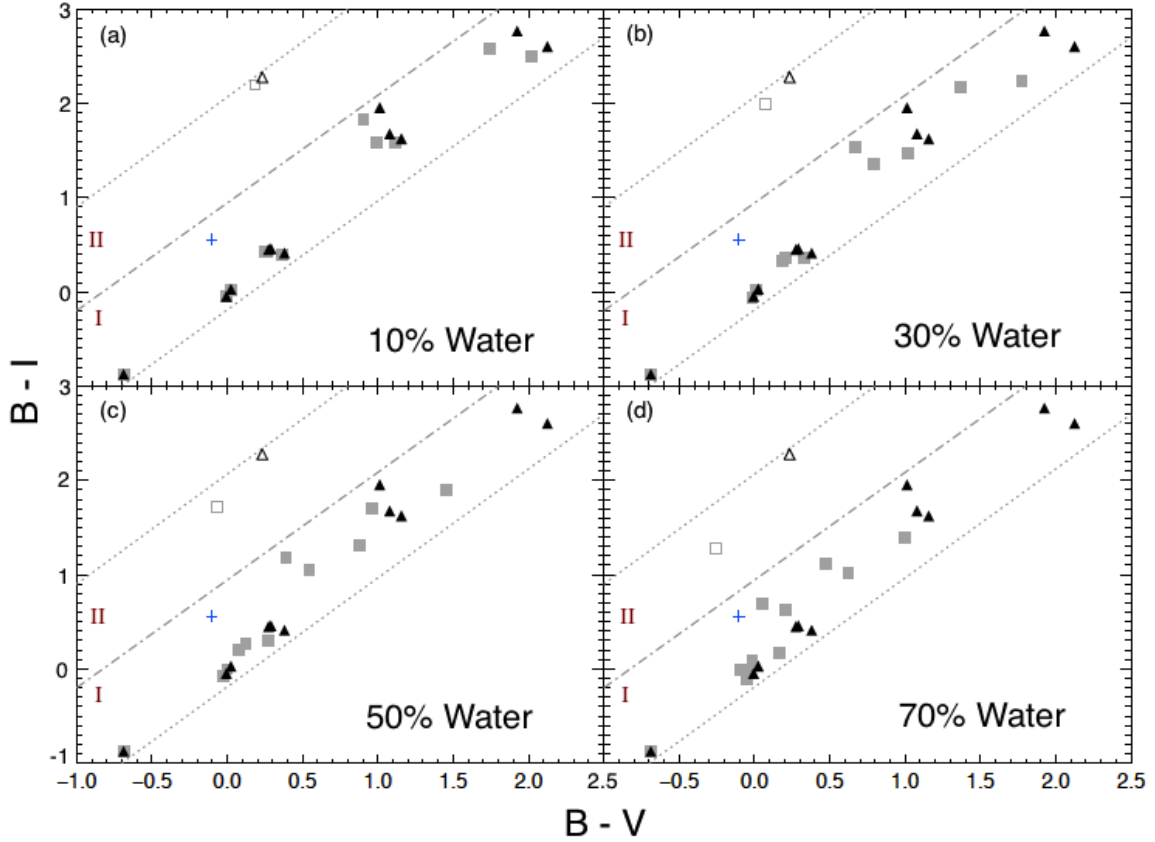

FIG. 4: **Color-color diagrams (as above) and water.**

> Filled triangles represent a planet completely covered by a particular surface. Filled squares denote the case when the planet is (a) 90%, (b) 70%, (c) 50%, and (d) 30% covered by a particular surface with the rest being liquid water. Trees are shown as unfilled triangles (complete coverage) and squares as reference to other VRE studies. The blue data point represents present-day Earth. Region I defines the area of extreme Earth surfaces; region II includes surface vegetation for non-extreme forms of life.

The VRE, indicated by the albedo of trees here, does not group with the *Synechococcus*-bearing bacterial mats and lichens, even though all three undergo photosynthesis and contain chlorophyll pigments. One explanation is that the bacterial mat used here is covered with a thin layer of water; therefore, the reflection spectrum has strong water absorption features longward of 0.70 µm. Lichens, on the other hand, are composite organisms composed of a fungus with a photosynthetic partner, usually either a green alga or a cyanobacterium. Therefore, properties of lichens are very different from those of isolated fungus or alga in culture (Kranner et al., 2005). Hence, the VRE signature in lichens is observed to be weak or sloping (Kiang et al., 2007a).

The albedo and therefore the position in a color-color diagram of vegetation or any chlorophyll-bearing photosynthetic organism on an extrasolar rocky planet depend on the radiation received from the host star. For example, the chlorophyll signature for planets around hot stars may have a "blue-edge" that reflects some of the high-energy radiation and prevents the leaves from overheating (Kiang et al., 2007b). The chlorophyll signature for planets orbiting cooler stars may appear black due to the total absorption of energy in the entire visible waveband such that plants gain as much available light as possible for photosynthetic metabolism (O'Malley-James et al., 2012). Therefore, the positions of trees, microbial mats, and lichens in Fig. 2 are only valid for an Earth analog planet orbiting around a Sun-like star and should be taken as guides. The albedo of vegetation and chlorophyll-bearing organisms for non-Sun-like stars requires further study.

Radiation is a major factor when considering the habitability on extrasolar planets since it is capable of causing extensive damage to nucleic acids, proteins, and lipids in the UV regime (Rothschild, 2009). For the case of present-day Earth, the presence of an ozone layer in the atmosphere acts as an envelope by absorbing UV radiation and thereby shields surface life from high doses of UV light. Such an ozone layer was absent on early Earth until about 2.3 billion years ago (Kasting and Catling, 2003), but life still remained abundant. In addition, the young Sun emitted substantially more UV radiation than its current values (Canuto et al., 1982; Cockell, 2000), making the surface of Earth extremely hazardous to most current life-forms that evolved in a low-radiation environment. Subsurface or ocean life would not have been affected by such radiation. Therefore, although radiation extremophiles are not yet found on Earth, we explore radiation as one of the physical extremes while looking for potential niches on extrasolar rocky planets.

Most of the extremophiles considered in this work are based on individual physical or geochemical extremes (e.g., temperature, pH). The strategy implemented in this paper is also applicable to "polyextremophiles," that are, organisms thriving in multiple environmental extremes as well as organisms found in new niches once their characteristics become available.

The results presented here were obtained by using the standard Johnson-Cousins broadband filters. Varying the bandpass definitions by using custom filters does not improve the results significantly. For further in-depth characterization, the bandpass definitions could be optimized to distinguish specific surfaces; or narrow-band filters could be used, provided a high enough SNR is available.

## 6. Conclusions

For direct imaging of discovered exoplanets, information on habitability can be explored by using atmospheric and surface properties of the planet as seen through the observations of Earth from interplanetary spacecrafts (e.g., Sagan et al., 1993; Geissler et al., 1995) and from atmospheric modeling studies (see, e.g., Des Marais et al., 2002; Traub and Jucks, 2002; Segura et al., 2003, 2007; Selsis et al., 2007; Kaltenegger et al., 2010, for an in-depth discussion). The low SNRs that are presently achievable limit high-resolution spectroscopic measurements of rocky exoplanets.

This paper shows how filter photometry can serve as a first step in the characterization of Earth analog exoplanets with different surfaces. We use a simple low-resolution color-color diagram to remotely characterize different types of rocky planet environments. We link those remotely detectable surface signatures to extreme forms of life that such environments could potentially support for aerobic as well as anaerobic atmospheres.

Our approach can be used to prioritize exoplanets for follow-up spectroscopy. An Earth-type rocky planet placed outside the contour region in Fig. 4 should receive lower priority for follow-up since the surface environment would not correspond to known environments on Earth supporting life based on our current knowledge. New discoveries of extreme life-forms could expand these contours in the future. Our results also indicate that the priority of the target planet for follow-up characterization should increase toward the lower left corner in the color-color diagram due to a higher probability of liquid water being indicated on the surface.

Detailed spectroscopic studies will be needed to learn more as to the potential habitability of extrasolar rocky planets.

## Acknowledgements

The authors would like to thank Dorian S. Abbot and the referee for comments that strengthened the paper as well as Lynn J. Rothschild for useful discussion on extremophiles. The authors acknowledge support from DFG funding ENP Ka 3142/1-1. Lisa Kaltenegger gratefully acknowledges support from the NASA Astrobiology Institute (NAI). Siddharth Hegde acknowledges support from the International Max Planck Research School for Astronomy and Cosmic Physics at the University of Heidelberg (IMPRS-HD), of which he is a fellow.

## Author Disclosure Statement

No competing financial interests exist.

## Abbreviations

AMD, acid mine drainage; SNR, signal-to-noise ratio; VRE, vegetation red edge.